\documentclass[pra, aps, showpacs,reprint]{revtex4-1}
\usepackage{boxedminipage}
\usepackage[utf8]{inputenc}
\usepackage{graphicx}
\usepackage{amsmath}
\usepackage{amssymb}
\usepackage{braket}
\usepackage{color}
\usepackage{subfigure}
\usepackage{tikz}
\usepackage{multibib}
\renewcommand\vec[1]{\ensuremath\boldsymbol{#1}}

\newcommand{\spinup}{\uparrow}
\newcommand{\spindown}{\downarrow}

\newcommand{\nearnb}[2]{\left< #1 , #2 \right>}

\newcommand{\Ins}[0]{\textnormal{AF}}

\newcommand{\Cond}[0]{N}

\newcommand{\Fin}[0]{f}
\newcommand{\Ini}[0]{i}

\DeclareMathOperator{\Tr}{Tr}

\usepackage[colorlinks=true,urlcolor=blue,menucolor=blue,citecolor=blue,linkcolor=blue]{hyperref}                                

\begin{document}

\title{Influence of disorder at Insulator-Metal interface on spin transport }

\author{Mahsa Seyed Heydari}
\author{Wolfgang Belzig}
\author{Niklas Rohling}
\affiliation{Department of Physics, University of Konstanz, Konstanz, Germany}

\date{March 28th 2024}

\begin{abstract}
Motivated by experimental work showing enhancement of spin transport between Yttrium Iron Garnet and Platinum by a thin antiferromagnetic insulator between them, we consider spin transport through the interface of a non-magnetic metal and compensated antiferromagnetically ordered insulator and focus on the significance of the interface itself.
The spin transport is carried by spin-polarized electrons in the metal and by magnons in the insulator.
We compute the spin current  in the presence of a spin accumulation in the metal, cause by the spin Hall effect, and a thermal gradient using Fermi’s Golden Rule in the presence of interfacial disorder.
For a perfectly clean interface, the in-plane momentum is conserved by the electron-magnon scattering events that govern the spin transport through the interface.
We calculate how disorder-induced broadening of scattering matrix elements with respect to the in-plane momentum influences the spin current. 
As a general result, we observe that for many experimental setups, specifically for high temperatures, one should expect a rather small effect of interface disorder on the measured spin current, while for small temperatures there is a significant reduction of a spin current with increasing disorder.

\end{abstract}

\maketitle
\section{Introduction}
Spintronics based on antiferromagnets (AFM) is a rising field due to its possible applications related to certain advantages compared to ferromagnetic spintronics. Specifically, the antiferromagnetic order is to large extends unaffected by magnetic fields and the high frequencies of its elementary excitations \cite{jungwirth2016antiferromagnetic,gomonay2018antiferromagnetic}.
For spin transport between metals and magnetically ordered insulators, there have been exciting results for the spin current through a thin antiferromagnetic insulator.
Namely, in trilayer systems consisting of a heavy metal, the thin antiferromagnetic layer, and an insulating ferromagnetic (FM) a spin-current enhancement was observed compared to the metal-FM bilayer system \cite{Wang:prl2014,Wang:prb2015,Lin2016enhancement}.
In general, the spin current in these systems can be generated by
(i) an electric current in the heavy metal parallel to the interface via the spin Hall effect effectively transferring angular momentum to the magnetically ordered insulator (spin transfer torque),
by (ii) a microwave field exciting spin waves in the magnetically ordered insulator (spin pumping), or by
(iii) a thermal gradient (spin Seebeck effect).
Independent of its generation, the nature of the spin current changes at the interface between the metal and the insulator from spin polarized electrons in the metal to spin waves in the insulator.

The spin transport through the interface between a non-magnetic metal and a ferromagnetically, antiferromagnetically or ferrimagnetically ordered insulator has been theoretically described in multiple papers \cite{bender2012electronic, kapelrud2013spin, Cheng:prl2014, PhysRevLett.108.246601, fjaerbu2017electrically, kamra:prl2017, zheng:prb2017, guemard:prb2022}.
The theoretical descriptions include
scattering theory with classical equations of motion for the sublattice magnetization \cite{Cheng:prl2014},
Fermi's Golden Rule treating the interface exchange coupling as a perturbation \cite{bender2012electronic,fjaerbu2017electrically}, and
Green's functions formalism applied to both, magnons and electrons \cite{guemard:prb2022}.

We note that the specifications of the interface between the metal and the antiferromagnetic insulator can be of crucial importance as they might be responsible for the sample dependence observed in experiments e.g. in Ref.~\cite{Wang:prl2014}.
Furthermore, the sign of the spin current between an antiferromagnetic insulator with two sublattices with opposite spin orientation depends on which of the sublattices couples more strongly to the metal \cite{Rohling:prb2023,tang2023spin}.

Here we focus on the influence of interface disorder on spin-transport.
So far in the theoretical description, either the clean limit with conservation of the components of the quasi momentum parallel to the interface was considered \cite{fjaerbu2017electrically,Rohling:prb2023} or the dirty limit where the scattering amplitudes are considered to be independent of the electron and magnon momenta \cite{guemard:prb2022}.
We present a general approach to investigate the influence of interface-roughness on the spin transport.

The remainder of the paper is organized as follows.
In Sec.~\ref{sec:dynamics}, we introduce the model for the metal, the antiferromagnetically ordered insulator and its exitations, i.e., magnons, and the interaction.
Then, in Sec.~\ref{sec:rates}, we compute the spin current caused by a spin-Hall-effect-induced spin accumulation under the influence of roughness-induced broadening of the scattering amplitudes. We follow the formalism using Fermi's Golden Rule from Refs.~ \cite{bender2012electronic,fjaerbu2017electrically,bender2014dynamic}.
We present our numerical results in Sec.~\ref{sec:res} finding most notably that the interface disorder leads to a reduction of the spin current most prominently for small temperatures.
Finally, we conclude in Sec.~\ref{sec:conclusions}. 
\section{Model}
\label{sec:dynamics}
\begin{figure}
\centering\includegraphics[width=\linewidth]{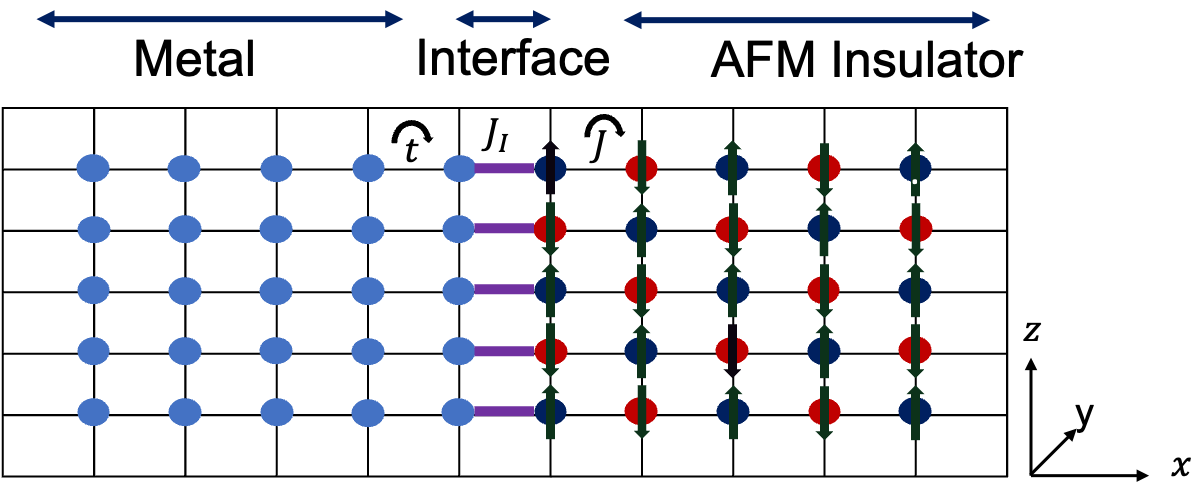}
\caption{Schematic representation of the lattice structure of the system including metal, interface and compensated AFM insulator.}
\label{Model1}
\end{figure}
The system under investigation is a bilayer consisting of a magnetically ordered insulator adjoining a nonmagnetic metal including a metal-insulator interface.
The lattice structure of the normal metal and the antiferromagnetic insulator is identical and cubic and the easy axis of the antiferromagnet is in $z$ direction considering the anisotropy term in the Hamiltonian.
As a consequence, the spins are pointing in $\pm z$ direction in the magnetically ordered ground state.
The $x$ direction is normal to the interface.
At the interface, there can be a spin accumulation on the metallic side cause by an electric current parallel to the interface via the spin Hall effect.
In metal-based spintronics, a spin accumulation is synonymous with a chemical potential difference between spin up and down bands.

In order to have spin current through the interface, there are different methods, such as
the spin Seebeck effect \cite{ashcroft2022solid, adachi2013theory} in which $T_M$ is magnon temperature and $T_N $ is normal metal temperature, where $T_{N} \neq T_{M}$,
spin transfer torque \cite{slonczewski1996current}  where $\Delta \mu \neq 0$ and
spin pumping where spin-waves in the insulator are excited by a microwave field \cite{tserkovnyak2002spin}.
Without considering external perturbations, the Hamiltonian includes three parts
\begin{equation}
    H= H_{N} + H_{M} + H_{I}.
\end{equation}
The first term describes electrons in the normal metal by using a tight-binding Hamiltonian that describes the energy of an electron in a simple cubic lattice by taking into account the hopping of electrons between neighboring lattice sites,
\begin{equation}
    H_{N} = -t \sum_{\langle i , j \rangle}\sum_{\sigma=\uparrow,\downarrow} \Big( c^\dagger_{i , \sigma} c_{j , \sigma} + c^\dagger_{j , \sigma} c_{i , \sigma}\Big).
    \label{eq:dynamics.metalHamiltonian}
\end{equation}
Here, $c^\dagger_{i,\sigma}$ and $c_{i,\sigma}$ are the creation and annihilation operators for an electron at site $i$ with spin $\sigma$, respectively. The expression $\langle i , j \rangle$ denotes nearest neighbors.
Including an easy-axis anisotropy term, the antiferromagnetic Hamiltonian can be described using the Heisenberg model,
\begin{equation}
		H_{M} = J \sum_{\nearnb{i}{j} \mid i,j \in \Ins} \vec{S}_i \cdot \vec{S}_j - K_z \sum_{i} S_{i z}^2 .
	\label{eq:dynamics.magnetHamiltonian}
\end{equation} 
Here, $S_i$ and $S_j$ are the spin operators at lattice sites $i$ and $j$, respectively and $J$ is the exchange interaction. In the second term $K_z$ is the anisotropy constant in $z$ direction which is preferred direction for the spins to alignment in their future.
Antiferromagnet insulator and metal are coupled at the interface. Noting that the interaction occurs only between the nearest neighbors across the interface the local exchange interaction between them is described as
\begin{equation}
		\tilde H_{I} = J_{I} \sum_{\nearnb{i}{j} \mid i \in \Ins, j \in \Cond} \vec{S}_i \cdot c^\dagger_{\sigma j}\boldsymbol\tau_{\sigma\sigma'}c_{\sigma' j} ,
	\label{eq:dynamics.InteractionHamiltonian}
\end{equation}
where $J_I$ is the exchange interaction between metal and AFM and $\boldsymbol\tau$ is a vector of Pauli matrices.

For the purpose of describing spin operators in terms of bosonic annihilation (creation) operators $a^{(\dagger)}$ and $b^{(\dagger)}$ the second-order truncated Holstein-Primakoff transformation,
$$S_{j+}=\hbar\sqrt{2s} a_j,~~S_{j-}=\hbar\sqrt{2s}a_j^\dagger, ~~S_{jz}=\hbar(s-a_j^\dagger a_j),$$
for the lattice site $j$ being part of sublattice A and
$$S_{j+}=\hbar\sqrt{2s} b^\dagger_j,~~S_{j-}=\hbar\sqrt{2s}b_j, ~~S_{jz}=\hbar(b_j^\dagger b_j-s)$$
for the lattice site $j$ being part of sublattice B
with $S_{j\pm}=S_{jx}\pm iS_{jy}$ is used.
With the assumption that the contributions from higher-order terms are negligible as well as the effect of the zeroth-order staggered field at the interface, we obtain an effective interaction Hamiltonian

 \begin{align}
     H_I  = &\frac{J_I \hbar^2}{2} \sum_{\langle i,j\rangle | i \in A , j \in N} \left( a_i c^\dagger_{\downarrow j} c_{\uparrow j} + a^\dagger _i c^\dagger_{\uparrow j} c_{\downarrow j}\right) \notag
     \\ & + \frac{J_I \hbar ^2}{2} \sum_{\langle i,j\rangle | i \in B , j \in N} \left( b^\dagger _i c^\dagger_{\downarrow j} c_{\uparrow j} + b_i c^\dagger_{\uparrow j} c_{\downarrow j} \right).
     \label{eq:scattering.coefficients.InteractionHamiltonian}
 \end{align}

\subsection{Scattering amplitude}
\begin{figure}[ht!]
\centering\includegraphics[width=0.7\linewidth]{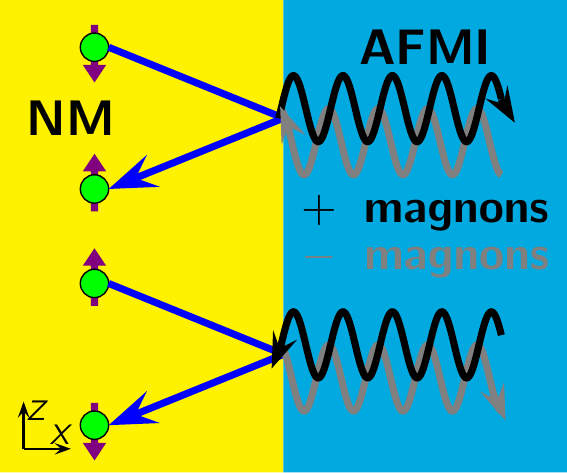}
\caption{Schematic representation of the system and the interfacial electron-magnon scattering. Two types of magnons, denoted by $\pm$, with opposite angular momentum exist, which are related by a reflection in the $x-y$ plane.
An electron flipping its spin in the scattering  from up to down corresponds to the creation of a $-$ magnon (gray) or the annihilation of a $+$ magnon (black) and the other way around for the electron spin changing from down to up at scattering.}
\label{Model}
\end{figure}

Next, substituting the delocalized states for magnons and scattering states for conduction electrons we will have
\begin{equation}
    H_{I} = \sum_{ \vec q  \vec k  \vec k^\prime} [ V^+_{ \vec q  \vec k  \vec k^\prime} \alpha^+_{\vec q} c^\dagger _{\downarrow \vec k} c _{\uparrow \vec k} + V^-_{ \vec q  \vec k  \vec k^\prime} \alpha^-_{\vec q} c^\dagger _{\uparrow \vec k} c _{\downarrow \vec k} ] +h.c.,
    \label{eq:scattering.coefficients}
\end{equation}
Here, the $V_{\vec q\vec k\vec k'}^\pm$ represent matrix elements while $(\alpha^\pm)^{(\dagger)}$ denotes the magnon annihilation (creation) operator with spin $\hbar$ in $\mp z$ direction.
We have
\begin{equation}
	V^\pm_{\vec q\vec k\vec k^\prime} = \sqrt{\frac{s}{2}}J_I\sin(k_xa)\sin(k_x'a) 
u_{\vec q 0}  F(\vec q,\vec k ,\vec k^\prime)
\end{equation}
where $u_{\vec q 0}$ is the amplitude of a magnon state $(\alpha^\pm)$ at the interface, $a$ is the lattice constant, and $\sin(k_xa)$ and $\sin(k_x'a)$ come from the amplitudes of the electronic eigenstates of Eq.~(\ref{eq:dynamics.metalHamiltonian}), denoted by $c_{\sigma \vec k}$, at the interface.
We will use the function $F(\vec q,\vec k, \vec k^\prime)$ in the following to quantify the roughness of the interface.
The electron-magnon interaction at the interface is described by a processes in which an electron spin is flipped during scattering and a magnon is either created or annihilated, as is shown in Fig.~\ref{Model}.
 By using the following Ansatz the magnonic eigenstates are determined \cite{fjaerbu2017electrically}
\begin{align}
			&\alpha^{+}_{\vec{q}} &= \sum_{i \in A} \frac{ u_{\vec{q}}\left( x_i \right) }{\sqrt{N/2}} e^{-i \vec{q}_{\parallel} \cdot\vec{r}_{i \parallel}} a_{i} - \sum_{i \in B} \frac{v_{\vec{p}}\left( x_i \right)}{\sqrt{N/2}} e^{i  \vec{q}_{\parallel}\cdot \vec{r}_{i \parallel}} b^{\dag}_{i}, \notag \\
			&\alpha^{-}_{\vec{q}} &= \sum_{i \in B} \frac{ u_{\vec{q}}\left( x_i \right) }{\sqrt{N/2}} e^{- i \vec{q}_{\parallel}\cdot \vec{r}_{i \parallel}} b_{i} - \sum_{i \in A} \frac{v_{\vec{p}} \left( x_i \right) }{\sqrt{N/2}} e^{i \vec{q}_{\parallel}\cdot \vec{r}_{i \parallel} } a^{\dag}_{i}  \, ,
	\label{eq:scattering.magnons.def.creationannihilation}%
 	\end{align}
where we introduced the in-plane momenta $\vec{q}_\parallel=(q_y,q_z)$ and position $\vec{r}_\parallel=(y,z)$.
The magnon energy for both types of magnons, with spin $\mp\hbar$, is taken in the long-wavelength limit as
\begin{equation}
    \varepsilon_{\vec q} = \sqrt{(\hbar vq)^2+ E^2_0}
    \label{ME}
\end{equation}
with $v = \sqrt{3\hbar } Ja $ being the spin-wave velocity and $E_0 = \hbar ^2 \sqrt{6 J K_z}$ is the spin-wave gap including the small anisotropy. This long-wavelength assumption is justified since the temperature is assumed smaller than the N\'eel temperature, which has the same order of magnitude as the exchange energy $~2J$.
 With this we can express the magnon amplitude at the interface in the long-wavelength limit as \cite{fjaerbu2017electrically}
\begin{equation}
    u_{\vec q 0} = \sqrt{\frac{3\hbar^2J}{\varepsilon_{\vec q}}}.
    \label{U_Q}
\end{equation}
The dispersion relation for the conduction electrons reads
\begin{equation}
    E_{\vec k}= -2 t(\cos{(k_x a)}+\cos{(k_y a)}+\cos{(k_z a)}) + 6t.
    \label{EE}
\end{equation}
in which $6t$ is the Fermi energy and electronic states are used as follows
\begin{equation}
    c_{\sigma \vec k} \sim \sum_j \exp{(- i \vec{k}_\parallel \cdot \vec r_{j \parallel})} \sin(k_x  (x_j{-}x_I{-}a)) c_{\sigma j}
\end{equation}
where $x_I$ is the $x$ coordinate of the interfacial metal layer.

\section{Transport rate}
\label{sec:rates}
To find the transport rate, we use Fermi's Golden Rule \cite{bender2012electronic} which describes the probability of transition a quantum state from an initial to a final state under the influence of a weak perturbation.
Due to the loss of phase coherence within each subsystem, the overall density matrix describing the insulator $\hat{\rho} _M$ and conductor $\hat{\rho} _{N}$  can be written by a decoupled state as 
$\hat{\rho} _{tot} = \hat{\rho}_{M} \otimes \hat{\rho}_{N}$. 
In equilibrium, the magnons follow the Bose-Einstein distribution and electrons the Fermi-Dirac distribution.
By considering the interaction term the system is taken out of equilibrium.
Since the interaction at the interface is assumed to be slow compared to thermalization effects so the system will be in the quasi-equilibrium state.
This asserts 
\begin{equation}
    \operatorname{Tr}[ \hat{\rho}_N \hat{c}_{\sigma  \vec k } ^\dagger \hat{c}_{ \sigma ^\prime \vec k^\prime  }] = n_F \left(\beta_N(E_{\vec k} - \mu _\sigma) \right) \delta _{k k^\prime } \delta _{\sigma \sigma^ \prime}
\end{equation}
\begin{equation}
    \operatorname{Tr}[ \hat{\rho}_{M} (\hat{\alpha}_{\vec q }) ^\dagger \hat{\alpha}^\pm_{\vec q^\prime  }] =n_B\left( \beta_M (\varepsilon^\pm _{\vec q} - \mu _L ) \right) \delta _{q q^\prime },
\end{equation} 
where $n_F(x)=[\exp(x)+1]^{-1} $ and $n_B(x)=[\exp(x)-1]^{-1}$ are the Fermi-Dirac and Bose-Einstein distributions, respectively, $\mu_\sigma$ is the chemical potential of the electrons with spin $\sigma$, $\mu_L$ is the magnon chemical quasi potential \cite{flebus2019chemical}, which will be set to zero in the remainder of the paper, $\beta_N = (k_B T_N)^{-1}$ and $\beta_M = (k_B T_M)^{-1}$ are the inverse temperatures of the electrons and magnons, respectively, and $k_B$ is the Boltzmann constant.
The difference $\Delta\mu=(\mu_\uparrow-\mu_\downarrow)/2$ is the spin accumulation in the metal.
It is straightforward to calculate the spin current (per interfacial area) flowing into the insulator from the conductor in dependence of temperatures and chemical potentials.

By considering $H_I$ as perturbation term, Fermi's Golden Rule provides the transition rate
\begin{equation}
    I_{i \rightarrow f} =\frac{2 \pi}{\hbar}|\langle \psi_f|H_I|\psi_i\rangle|^2 \delta (E_f-E_i)
    \label{FG}
\end{equation}
such that $\psi_{ \Fin }$ and $\psi_{\Ini}$ are the final and initial state and $E_f$ and $E_i$ their respective energies of the non-interacting model.
We obtain for the individual magnon branches the transport rates
\begin{align}		
		I^{\pm} = \!\frac{2 \pi}{\hbar} |V^{\pm}_{\vec{q} \vec{k} \vec{k}'}|^2 &\!\!\sum_{\vec{q}\vec{k} \vec{k}'} 
 \left[ \Tr \left\{ \rho    \alpha^{\pm}_{\vec{q}} c^{\dag}_{\spindown \vec{k}} c_{\spinup \vec{k}'}  (\alpha^\pm_{\vec{q}})^\dag  c^{\dag}_{\spinup \vec{k}'} c_{\spindown \vec{k}}  \right\} \right.\notag \\
		- &\left.\!\!\Tr \left\{\! \rho  (\alpha^\pm_{\vec{q}})^\dag  c^{\dag}_{\spinup \vec{k}'} c_{\spindown \vec{k}}  \alpha^{\pm}_{\vec{q}} c^{\dag}_{\spindown \vec{k}} c_{\spinup \vec{k}'} \!\right\}\! \right] \!\delta \!\left( E_{ \Fin } {-} E_{ \Ini } \right) \!.
	\label{eq: rates.FermiCurrent}
\end{align}
Here, the creation and annihilation of magnons causes a change in the spin angular momentum of the itinerant electrons.
Note that magnon-magnon and electron-electron interactions in the bulk are only included in so far that we assume them to contribute, combined with coupling to phonons, to the thermalization.
This is consistent with the assumption of the magnon number is not affected by bulk effects.
Due to the symmetry of the Hamiltonian, the total angular momentum and total number of magnons of each type is conserved.
Another assumption is that the Fermi energy $E_F$ and the exchange energy $J$ are considerably larger than the magnon gap $E_0$ and the spin accumulation $\Delta\mu$.
Depending on the direction of the scattered spin, both types of magnons, those with spin $-\hbar$ and  $\hbar$  will be appeared and they will be affected differently by the polarization of spin accumulation.

\subsection{Current}
We are interested in calculating the overall rate transition of the magnon number $I$.
By starting from  Eq.~(\ref{FG})
and summing over all final states
to have the probability of scattering out of the initial state into the final states we obtain \cite{fjaerbu2017electrically}
\begin{align}		
		I^\pm =& \frac{2\pi}{\hbar}\int d\varepsilon
	[\varepsilon \pm\Delta\mu]
		\left[n_B\left(\beta _N(\varepsilon \pm \Delta\mu)\right) {-} n_B\!\left(\!\beta_M \varepsilon\!\right)\!\right] \notag \\
		& |V_{\vec q\vec k\vec k'}^\pm|^2 \sum_{\vec q\vec k\vec k'}
		\delta(E_F{-}E_{\vec k})
		\delta(\varepsilon{-}\varepsilon_{\vec q})
	\delta(\varepsilon_{\vec q}+E_{\vec k}{-}E_{\vec k^\prime})
 \label{I1}
\end{align}
where $ \pm $ again indicates two types of magnons that exist in the AFM.
Note that Eq.~(\ref{I1}) results from rewriting the Fermi-Dirac distributions for the electrons \cite{bender2012electronic}.
Eq.~(\ref{I1}) is simplified as 
\begin{align}		
		I^\pm =& \frac{2\pi}{\hbar}\int d\vec q d\vec k d\vec k'\  |V_{\vec q\vec k\vec k'}^\pm|^2 
	(\varepsilon_{\vec q} \pm \Delta\mu) \delta(E_F{-}E_{\vec k}) \notag \\
&\times\left[n_B\left(\beta _N(\varepsilon_{\vec q} \pm \Delta\mu)\right) {-} n_B\!\left(\!\beta_M \varepsilon_{\vec q}\!\right)\!\right]
		\delta(\varepsilon_{\vec q}{+}E_{\vec k}{-}E_{\vec k'})
  \label{CU}
\end{align}
Two cases are investigated here, in the case of clean limit, there is a conservation of in-plane quasi momentum and by adding disorder at the interface, the conservation is broken and the interface is not considered as clean anymore.
We quantify this effect via the function $F({\vec q,\vec k ,\vec k^\prime})$
\begin{equation}
   F({\vec q,\vec k ,\vec k^\prime})=
\begin{cases}
 \delta _{\vec k_{\parallel} + \vec q_{\parallel} , \vec k^\prime _{\parallel}}  & \text{for the clean limit } \\
B_\sigma (\vec Q) & \text{with disorder } 
\end{cases}
\end{equation}
introducing $\vec Q=Q_y\vec e_y +Q_y\vec e_z$ with
$Q_z = q_z-k_z+k_z^\prime$ and $Q_y = q_y-k_y+k_y^\prime$.
Note that we include only Umklapp processes in the scattering matrix elements $V^\pm_{\vec q\vec k\vec k'}$ as those processes are dominating for $J_I/t\ll1$\cite{fjaerbu2017electrically}.
The Gaussian distribution function is used in order to describe the effect of disorder 
\begin{equation}
B_\sigma (\vec Q) =
     \frac{1}{2 \pi\sigma^2 }\exp{\Big(-\frac{\vec Q^2}{2\sigma^2}\Big)}.
     \label{disorder}
  \end{equation}
To avoid artefacts from a specific form of the function $B_\sigma (\vec Q)$, we also performed the computation with a Lorentzian distribution function.
By using the dispersion relation for the conduction electrons in Eq.~(\ref{EE}), the integral over $k_x$ is calculated. 
The system is considered exactly at half filling and the Fermi energy is $6t$.
The spin transport is dominated by electrons at energies close to the Fermi energy $E\approx E_F$.
Under this assumption one approximation is considered: In the term $\delta(\varepsilon_{\vec q}^\pm{+}E_{\vec k}{-}E_{\vec k'})$ the magnon energy $\varepsilon_{\vec q}$ can be disregarded due to the fact that it is small compare to the $ E_{\vec k}$.
We obtain $\delta(E_{F}{-}E_{\vec k'})   = \frac{-1}{2t} \delta (\cos{k'_x a}+ \cos{k'_y a}+\cos{k'_z a}) $.
We see that the approximation introduced above, accounts for dynamics near the Fermi surface.
Now, we solve integral over $k^\prime_x$.
In order to write the equation in more compressed way, we introduce
\begin{align}
     f(k_y,k_z)= &\sqrt{1{-}(\cos(k_ya) {+} \cos(k_za))^2 }\notag\\
                  &\times\theta(1-|\cos(k_ya) {+} \cos( k_za)|)
\end{align}

\subsection{Numerical solution for the current integral} 
By using Eq.~(\ref{CU}) and Eq.~(\ref{disorder}),  a seven-fold integral remains and  should be solved.
It can be simplified by rotating the coordinate system by replacing $k_y$ and $k_z$ with average and difference as of $\overline{k}_{z,y}$ and $\Delta k_{z,y}$, in order to separate dependency of the integration in magnonic and electronic momentum so new axes are introduced as follows
$$\bar k_z = \frac{ k^\prime_z +  k_z}{2} ,~~~ \bar k_y = \frac{ k^\prime_y +  k_y}{2},$$
$$ \Delta k_z = { k^\prime_z-  k_z}  ,~~~ \Delta k_y = { k^\prime_y-  k_y}, $$
\begin{equation}
    f(\vec k_{\parallel},\vec k^\prime_{\parallel}) = \bar{f} (\bar{\vec k}_\parallel , \Delta \vec k_{\parallel} ).
\end{equation}
We also define
\begin{equation}
     h(\vec q)=\frac{\varepsilon_{\vec q}\pm  \Delta\mu}{\varepsilon_{\vec q}}  
	\left[n_B\left(\beta _N (\varepsilon_{\vec q}{\pm}\Delta\mu \right)) - n_B\left( \beta_M\varepsilon_{\vec q} \right)\right]
\end{equation}
and
rewrite Eq.~(\ref{CU}) as
\begin{equation}
I^\pm  \sim   \int h(\vec q)  \bar{f} (\bar{\vec k}_\parallel , \Delta \vec k_{\parallel} )
 B_\sigma(\vec q , \Delta k_{\parallel}) d\bar{\vec k}_\parallel d(\Delta \vec k_{\parallel}) d\vec q.
 \label{CU2}
\end{equation}
The function $\bar{f} (\bar{\vec k}_\parallel , \Delta \vec k_{\parallel} )$ can be integrated over $\bar{\vec k}_\parallel$ separately since there is no dependency on $\vec q$ and leads us to define
\begin{equation}
\int  \bar f(\bar{ \vec k}_{\parallel} , \Delta\vec k_{\parallel}) d \bar{\vec  k}_{\parallel}=  \bar F(\Delta k_{\parallel} ).
\end{equation}
We solve this integral  numerically.
\begin{figure*}[t!]
\centering        
\subfigure[]{\label{FIG1a}\includegraphics[width=.48\textwidth]{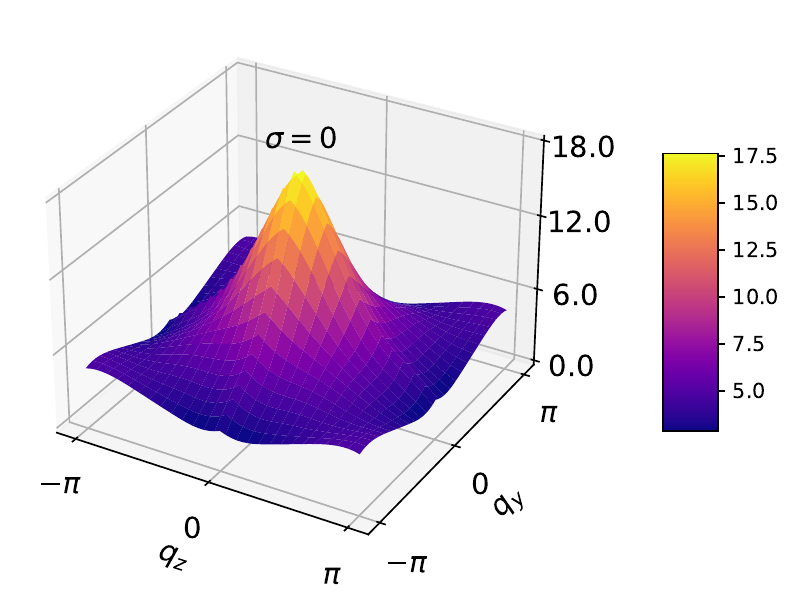}}
\subfigure[]{\label{FIG1b}
\includegraphics[width=.48\textwidth]{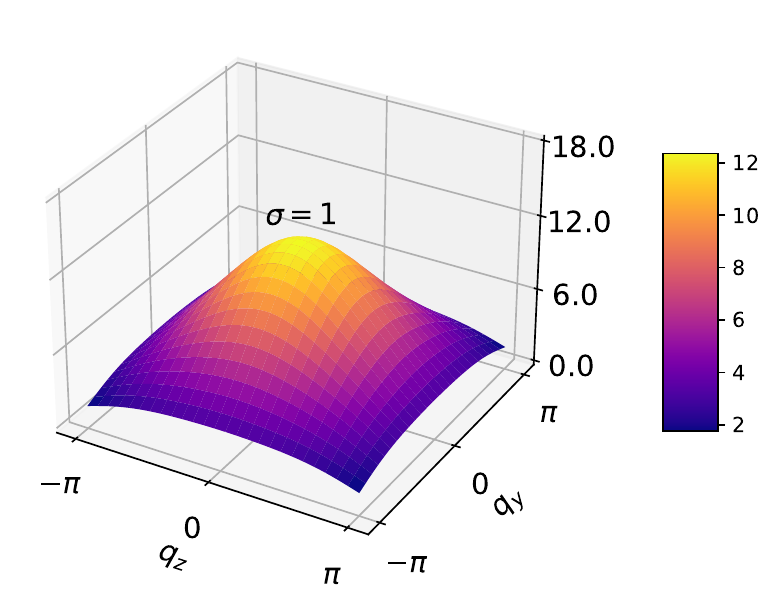}}
\caption{(a) The function  $\tilde{F}(q_y,q_z)$, defined in Eq.~(\ref{ft}),  in the clean limit $\sigma = 0$.
(b) $\tilde{F}(q_y,q_z)$ for disordered interface at $\sigma = 1 $ using a Gaussian distribution function. }
\label{fig2}
\end{figure*}
Next, we will have a 2D convolution integration as follows
\begin{equation} 
 \tilde{ F}(\vec q_{\parallel}) =  
 \int B_\sigma(\vec q_{\parallel}-\Delta \vec k_{\parallel}  )\bar F(\Delta \vec k_{\parallel} ) d(\Delta \vec k_\parallel)
 \label{ft}
 \end{equation}
Then, the remaining integrals are finally
\begin{equation}
    I^\pm \sim \int \tilde{ F}(\vec q_{\parallel}) h(\vec q) d\vec q.
\end{equation}

In Fig.~\ref{fig2}, Eq.~(\ref{ft}) is plotted in the case of clean limit, $\tilde F_{\rm clean}(\vec q_\parallel) = \bar F(\vec q_\parallel)$ and
in the presence of disorder.
It can be seen how broadening of the function $\tilde F(\vec q_\parallel)$ happens from the Gaussian replacing the $\delta$ peak in the clean interface shown in Fig \ref{FIG1b},
also it is visible that as an effect of disorder the peak goes lower.

\section{Spin Current in AFM}
\label{sec:res}
In order to evaluate the effect of the spin accumulation, the Bose-Einstein distribution function and Gaussian distribution function included in the formula for the current Eq.~(\ref{q1}) have to  be considered.
From
\begin{equation}
    I^{\pm} \!=\! 
    \int\! \frac{\varepsilon_{\vec q}\pm \Delta\mu}{\varepsilon_{\vec q}}
	\left[n_B\left(\beta _N (\varepsilon_{\vec q}{\pm}\Delta\mu \right)) {-} n_B\left( \varepsilon_{\vec q} \beta_M\right)\right]
 \tilde{F}(\vec q_\parallel) d \vec q,
 \label{q1}
	\end{equation}
we obtain the spin current as 
 $I_{\sigma} = I^--I^+$.
 \begin{figure*}
\centering        
\subfigure[]{\label{FIGa}\includegraphics[width=.45\textwidth]{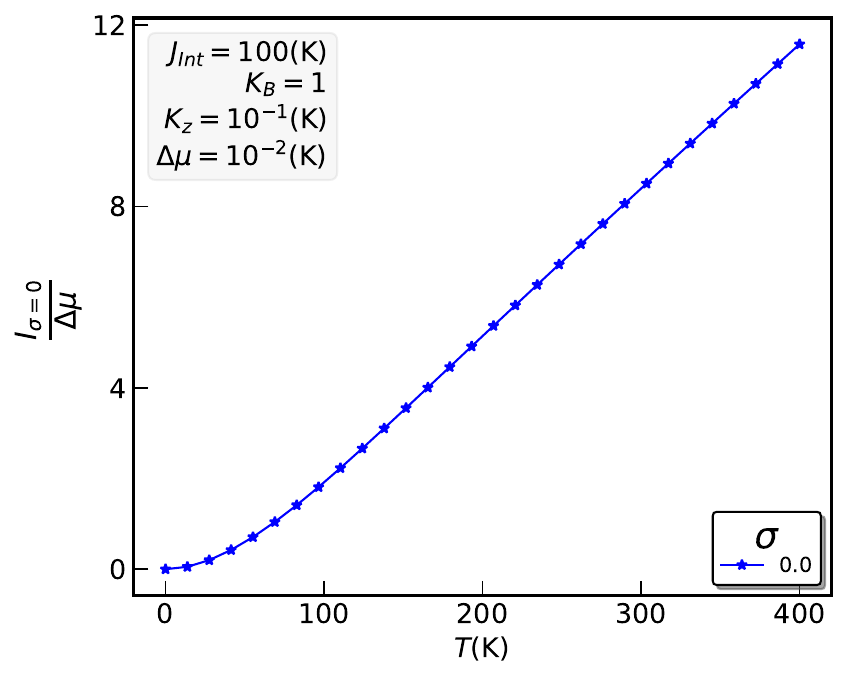}}
\subfigure[]{\label{FIGb}
\includegraphics[width=.47\textwidth]{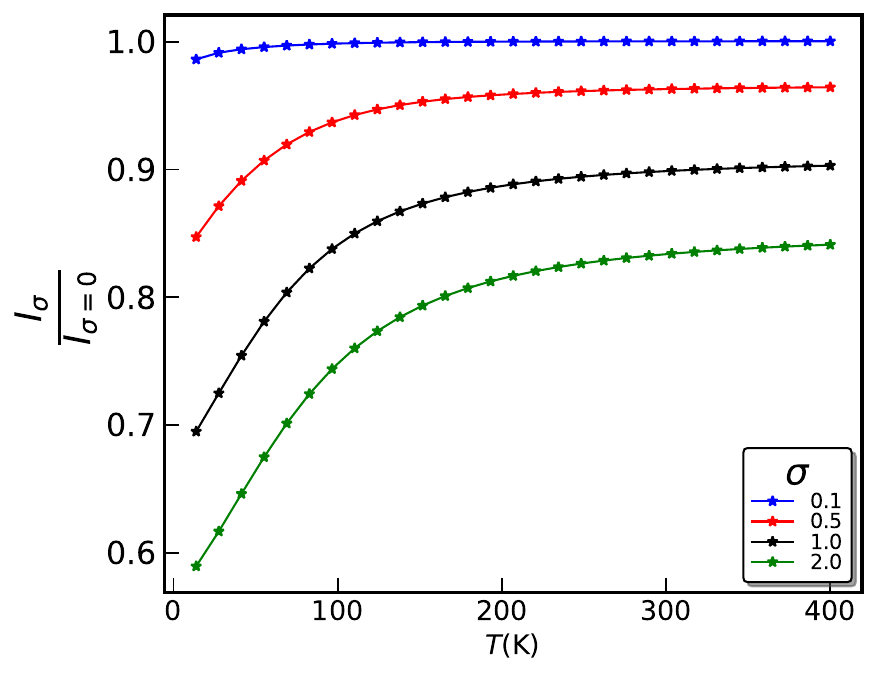}}
\caption{ (a) Spin current $I_{\sigma{=}0}$ as function of temperature in the clean interface limit.
(b) Ratio of spin current $I_\sigma$ with Gaussian broadening of $\tilde F(\vec q_\parallel)$ to $I_{\sigma{=}0}$ as function of temperature for four strengths of disorder.}
\label{fig31}
\end{figure*}
The index $\sigma$ indicates the $\sigma$ dependence via the function  $\tilde F(\vec q_\parallel)$.
The result for the ratio of the current with respect to the current of the clean interface, $I_{\sigma{=}0}$, as a function of temperature for various strengths of broadening
 is shown in Fig \ref{fig31}.
It can be seen that by increasing the disorder, i.e., increasing values of $\sigma$ the current passing through the antiferromagnetic insulator-metal interface decreases.
Specifically, this effect is remarkable for low temperature.
We found similar results for the Lorentzian form of $B_\sigma$.

\subsection{Expansion of spin accumulation}
In order to check the linearity of the current with respect to $\Delta \mu$, the formula for the spin current, Eq.~(\ref{q1}), is expanded in $\Delta \mu$ and $\Delta T$ for $\Delta\mu,\Delta T\ll T_N$.
We obtain
\begin{align}
    I_\sigma \sim 
        2\Delta\mu\beta_N n_B^\prime(\varepsilon_{\vec q} \beta _N) \left( 1+ \frac{\Delta T}{T_N} \right) 
\end{align}
where $n_B'(x)=dn_B(x)/dx$.
The current at $\Delta \mu \rightarrow 0$ is linear and, due to compensated interface, the temperature gradient alone does not cause spin current.

\section{Conclusions}
\label{sec:conclusions}
Following experiments highlighting the possibilities of a thin antiferromagnetic insulators for spintronics applications \cite{Wang:prb2015,Wang:prl2014,Lin2016enhancement}, we focused here on the importance of the interface properties which naturally appears to be more significant for thin layers.
A spin current is produced by a spin accumulation in the metal.
Our study covered the case of a clean interface as well as the influence of disorder via the broadening of the scattering matrix elements with respect to the in-plane momentum quantified by using Gaussian or Lorentzian distribution.
By increasing the effect of disorder introduced the quantity as $\sigma$, the spin current is in general reduced.
Moreover, we found that at low temperatures, the reduction of the spin current can be much more prominent.
Future work might cover also influence of disorder on the electronic states as well as disorder between magnetically ordered insulators in a trilayer structure.

\section*{Acknowledgements}
N.R. was financially supported by the German Research Foundation (DFG) project ID~417034116.
W.B. and N.R. acknowledge support from the Excellence Strategy at the University of Konstanz via the Free Space for Creativity program.
M.S.H. and N.R. acknowledge support via the Research Seed Capital (RiSC) funding scheme by the Ministry of Science, Research and Arts Baden W\"urttemberg as well as by Research Initiative programme within the Excellence Strategy at the University of Konstanz.

\section*{Author contributions}
N.R. developed the idea to investigate the influence of interfacial disorder on spin transport.
W.B. suggested the model for the effect of interfacial disorder.
M.S.H. developed the model and performed the computations under supervision of N.R. and W.B.
All authors participated in the discussion of the results and in writing the manuscript.

\section*{Competing interests}
The authors declare no competing interests.


\begin{thebibliography}{21}%
\makeatletter
\providecommand \@ifxundefined [1]{%
 \@ifx{#1\undefined}
}%
\providecommand \@ifnum [1]{%
 \ifnum #1\expandafter \@firstoftwo
 \else \expandafter \@secondoftwo
 \fi
}%
\providecommand \@ifx [1]{%
 \ifx #1\expandafter \@firstoftwo
 \else \expandafter \@secondoftwo
 \fi
}%
\providecommand \natexlab [1]{#1}%
\providecommand \enquote  [1]{``#1''}%
\providecommand \bibnamefont  [1]{#1}%
\providecommand \bibfnamefont [1]{#1}%
\providecommand \citenamefont [1]{#1}%
\providecommand \href@noop [0]{\@secondoftwo}%
\providecommand \href [0]{\begingroup \@sanitize@url \@href}%
\providecommand \@href[1]{\@@startlink{#1}\@@href}%
\providecommand \@@href[1]{\endgroup#1\@@endlink}%
\providecommand \@sanitize@url [0]{\catcode `\\12\catcode `\$12\catcode
  `\&12\catcode `\#12\catcode `\^12\catcode `\_12\catcode `\%12\relax}%
\providecommand \@@startlink[1]{}%
\providecommand \@@endlink[0]{}%
\providecommand \url  [0]{\begingroup\@sanitize@url \@url }%
\providecommand \@url [1]{\endgroup\@href {#1}{\urlprefix }}%
\providecommand \urlprefix  [0]{URL }%
\providecommand \Eprint [0]{\href }%
\providecommand \doibase [0]{http://dx.doi.org/}%
\providecommand \selectlanguage [0]{\@gobble}%
\providecommand \bibinfo  [0]{\@secondoftwo}%
\providecommand \bibfield  [0]{\@secondoftwo}%
\providecommand \translation [1]{[#1]}%
\providecommand \BibitemOpen [0]{}%
\providecommand \bibitemStop [0]{}%
\providecommand \bibitemNoStop [0]{.\EOS\space}%
\providecommand \EOS [0]{\spacefactor3000\relax}%
\providecommand \BibitemShut  [1]{\csname bibitem#1\endcsname}%
\let\auto@bib@innerbib\@empty
\bibitem [{\citenamefont {Jungwirth}\ \emph {et~al.}(2016)\citenamefont
  {Jungwirth}, \citenamefont {Marti}, \citenamefont {Wadley},\ and\
  \citenamefont {Wunderlich}}]{jungwirth2016antiferromagnetic}%
  \BibitemOpen
  \bibfield  {author} {\bibinfo {author} {\bibfnamefont {T.}~\bibnamefont
  {Jungwirth}}, \bibinfo {author} {\bibfnamefont {X.}~\bibnamefont {Marti}},
  \bibinfo {author} {\bibfnamefont {P.}~\bibnamefont {Wadley}}, \ and\ \bibinfo
  {author} {\bibfnamefont {J.}~\bibnamefont {Wunderlich}},\ }\href {\doibase
  10.1038/nnano.2016.18} {\bibfield  {journal} {\bibinfo  {journal} {Nature
  nanotechnology}\ }\textbf {\bibinfo {volume} {11}},\ \bibinfo {pages} {231}
  (\bibinfo {year} {2016})}\BibitemShut {NoStop}%
\bibitem [{\citenamefont {Gomonay}\ \emph {et~al.}(2018)\citenamefont
  {Gomonay}, \citenamefont {Baltz}, \citenamefont {Brataas},\ and\
  \citenamefont {Tserkovnyak}}]{gomonay2018antiferromagnetic}%
  \BibitemOpen
  \bibfield  {author} {\bibinfo {author} {\bibfnamefont {O.}~\bibnamefont
  {Gomonay}}, \bibinfo {author} {\bibfnamefont {V.}~\bibnamefont {Baltz}},
  \bibinfo {author} {\bibfnamefont {A.}~\bibnamefont {Brataas}}, \ and\
  \bibinfo {author} {\bibfnamefont {Y.}~\bibnamefont {Tserkovnyak}},\ }\href
  {\doibase 10.1038/s41567-018-0049-4} {\bibfield  {journal} {\bibinfo
  {journal} {Nature Physics}\ }\textbf {\bibinfo {volume} {14}},\ \bibinfo
  {pages} {213} (\bibinfo {year} {2018})}\BibitemShut {NoStop}%
\bibitem [{\citenamefont {Wang}\ \emph {et~al.}(2014)\citenamefont {Wang},
  \citenamefont {Du}, \citenamefont {Hammel},\ and\ \citenamefont
  {Yang}}]{Wang:prl2014}%
  \BibitemOpen
  \bibfield  {author} {\bibinfo {author} {\bibfnamefont {H.}~\bibnamefont
  {Wang}}, \bibinfo {author} {\bibfnamefont {C.}~\bibnamefont {Du}}, \bibinfo
  {author} {\bibfnamefont {P.~C.}\ \bibnamefont {Hammel}}, \ and\ \bibinfo
  {author} {\bibfnamefont {F.}~\bibnamefont {Yang}},\ }\href {\doibase
  10.1103/PhysRevLett.113.097202} {\bibfield  {journal} {\bibinfo  {journal}
  {Phys. Rev. Lett.}\ }\textbf {\bibinfo {volume} {113}},\ \bibinfo {pages}
  {097202} (\bibinfo {year} {2014})}\BibitemShut {NoStop}%
\bibitem [{\citenamefont {Wang}\ \emph {et~al.}(2015)\citenamefont {Wang},
  \citenamefont {Du}, \citenamefont {Hammel},\ and\ \citenamefont
  {Yang}}]{Wang:prb2015}%
  \BibitemOpen
  \bibfield  {author} {\bibinfo {author} {\bibfnamefont {H.}~\bibnamefont
  {Wang}}, \bibinfo {author} {\bibfnamefont {C.}~\bibnamefont {Du}}, \bibinfo
  {author} {\bibfnamefont {P.~C.}\ \bibnamefont {Hammel}}, \ and\ \bibinfo
  {author} {\bibfnamefont {F.}~\bibnamefont {Yang}},\ }\href {\doibase
  10.1103/PhysRevB.91.220410} {\bibfield  {journal} {\bibinfo  {journal} {Phys.
  Rev. B}\ }\textbf {\bibinfo {volume} {91}},\ \bibinfo {pages} {220410}
  (\bibinfo {year} {2015})}\BibitemShut {NoStop}%
\bibitem [{\citenamefont {Lin}\ \emph {et~al.}(2016)\citenamefont {Lin},
  \citenamefont {Chen}, \citenamefont {Zhang},\ and\ \citenamefont
  {Chien}}]{Lin2016enhancement}%
  \BibitemOpen
  \bibfield  {author} {\bibinfo {author} {\bibfnamefont {W.}~\bibnamefont
  {Lin}}, \bibinfo {author} {\bibfnamefont {K.}~\bibnamefont {Chen}}, \bibinfo
  {author} {\bibfnamefont {S.}~\bibnamefont {Zhang}}, \ and\ \bibinfo {author}
  {\bibfnamefont {C.~L.}\ \bibnamefont {Chien}},\ }\href {\doibase
  10.1103/PhysRevLett.116.186601} {\bibfield  {journal} {\bibinfo  {journal}
  {Phys. Rev. Lett.}\ }\textbf {\bibinfo {volume} {116}},\ \bibinfo {pages}
  {186601} (\bibinfo {year} {2016})}\BibitemShut {NoStop}%
\bibitem [{\citenamefont {Bender}\ \emph
  {et~al.}(2012{\natexlab{a}})\citenamefont {Bender}, \citenamefont {Duine},\
  and\ \citenamefont {Tserkovnyak}}]{bender2012electronic}%
  \BibitemOpen
  \bibfield  {author} {\bibinfo {author} {\bibfnamefont {S.~A.}\ \bibnamefont
  {Bender}}, \bibinfo {author} {\bibfnamefont {R.~A.}\ \bibnamefont {Duine}}, \
  and\ \bibinfo {author} {\bibfnamefont {Y.}~\bibnamefont {Tserkovnyak}},\
  }\href {\doibase 10.1103/PhysRevLett.108.246601} {\bibfield  {journal}
  {\bibinfo  {journal} {Phys. Rev. Lett.}\ }\textbf {\bibinfo {volume} {108}},\
  \bibinfo {pages} {246601} (\bibinfo {year} {2012}{\natexlab{a}})}\BibitemShut
  {NoStop}%
\bibitem [{\citenamefont {Kapelrud}\ and\ \citenamefont
  {Brataas}(2013)}]{kapelrud2013spin}%
  \BibitemOpen
  \bibfield  {author} {\bibinfo {author} {\bibfnamefont {A.}~\bibnamefont
  {Kapelrud}}\ and\ \bibinfo {author} {\bibfnamefont {A.}~\bibnamefont
  {Brataas}},\ }\href {\doibase 10.1103/PhysRevLett.111.097602} {\bibfield
  {journal} {\bibinfo  {journal} {Phys. Rev. Lett.}\ }\textbf {\bibinfo
  {volume} {111}},\ \bibinfo {pages} {097602} (\bibinfo {year}
  {2013})}\BibitemShut {NoStop}%
\bibitem [{\citenamefont {Cheng}\ \emph {et~al.}(2014)\citenamefont {Cheng},
  \citenamefont {Xiao}, \citenamefont {Niu},\ and\ \citenamefont
  {Brataas}}]{Cheng:prl2014}%
  \BibitemOpen
  \bibfield  {author} {\bibinfo {author} {\bibfnamefont {R.}~\bibnamefont
  {Cheng}}, \bibinfo {author} {\bibfnamefont {J.}~\bibnamefont {Xiao}},
  \bibinfo {author} {\bibfnamefont {Q.}~\bibnamefont {Niu}}, \ and\ \bibinfo
  {author} {\bibfnamefont {A.}~\bibnamefont {Brataas}},\ }\href {\doibase
  10.1103/PhysRevLett.113.057601} {\bibfield  {journal} {\bibinfo  {journal}
  {Phys. Rev. Lett.}\ }\textbf {\bibinfo {volume} {113}},\ \bibinfo {pages}
  {057601} (\bibinfo {year} {2014})}\BibitemShut {NoStop}%
\bibitem [{\citenamefont {Bender}\ \emph
  {et~al.}(2012{\natexlab{b}})\citenamefont {Bender}, \citenamefont {Duine},\
  and\ \citenamefont {Tserkovnyak}}]{PhysRevLett.108.246601}%
  \BibitemOpen
  \bibfield  {author} {\bibinfo {author} {\bibfnamefont {S.~A.}\ \bibnamefont
  {Bender}}, \bibinfo {author} {\bibfnamefont {R.~A.}\ \bibnamefont {Duine}}, \
  and\ \bibinfo {author} {\bibfnamefont {Y.}~\bibnamefont {Tserkovnyak}},\
  }\href {\doibase 10.1103/PhysRevLett.108.246601} {\bibfield  {journal}
  {\bibinfo  {journal} {Phys. Rev. Lett.}\ }\textbf {\bibinfo {volume} {108}},\
  \bibinfo {pages} {246601} (\bibinfo {year} {2012}{\natexlab{b}})}\BibitemShut
  {NoStop}%
\bibitem [{\citenamefont {Fj\ae{}rbu}\ \emph {et~al.}(2017)\citenamefont
  {Fj\ae{}rbu}, \citenamefont {Rohling},\ and\ \citenamefont
  {Brataas}}]{fjaerbu2017electrically}%
  \BibitemOpen
  \bibfield  {author} {\bibinfo {author} {\bibfnamefont {E.~L.}\ \bibnamefont
  {Fj\ae{}rbu}}, \bibinfo {author} {\bibfnamefont {N.}~\bibnamefont {Rohling}},
  \ and\ \bibinfo {author} {\bibfnamefont {A.}~\bibnamefont {Brataas}},\ }\href
  {\doibase 10.1103/PhysRevB.95.144408} {\bibfield  {journal} {\bibinfo
  {journal} {Phys. Rev. B}\ }\textbf {\bibinfo {volume} {95}},\ \bibinfo
  {pages} {144408} (\bibinfo {year} {2017})}\BibitemShut {NoStop}%
\bibitem [{\citenamefont {Kamra}\ and\ \citenamefont
  {Belzig}(2017)}]{kamra:prl2017}%
  \BibitemOpen
  \bibfield  {author} {\bibinfo {author} {\bibfnamefont {A.}~\bibnamefont
  {Kamra}}\ and\ \bibinfo {author} {\bibfnamefont {W.}~\bibnamefont {Belzig}},\
  }\href {\doibase 10.1103/PhysRevLett.119.197201} {\bibfield  {journal}
  {\bibinfo  {journal} {Phys. Rev. Lett.}\ }\textbf {\bibinfo {volume} {119}},\
  \bibinfo {pages} {197201} (\bibinfo {year} {2017})}\BibitemShut {NoStop}%
\bibitem [{\citenamefont {Zheng}\ \emph {et~al.}(2017)\citenamefont {Zheng},
  \citenamefont {Bender}, \citenamefont {Armaitis}, \citenamefont {Troncoso},\
  and\ \citenamefont {Duine}}]{zheng:prb2017}%
  \BibitemOpen
  \bibfield  {author} {\bibinfo {author} {\bibfnamefont {J.}~\bibnamefont
  {Zheng}}, \bibinfo {author} {\bibfnamefont {S.}~\bibnamefont {Bender}},
  \bibinfo {author} {\bibfnamefont {J.}~\bibnamefont {Armaitis}}, \bibinfo
  {author} {\bibfnamefont {R.~E.}\ \bibnamefont {Troncoso}}, \ and\ \bibinfo
  {author} {\bibfnamefont {R.~A.}\ \bibnamefont {Duine}},\ }\href {\doibase
  10.1103/PhysRevB.96.174422} {\bibfield  {journal} {\bibinfo  {journal} {Phys.
  Rev. B}\ }\textbf {\bibinfo {volume} {96}},\ \bibinfo {pages} {174422}
  (\bibinfo {year} {2017})}\BibitemShut {NoStop}%
\bibitem [{\citenamefont {Guemard}\ and\ \citenamefont
  {Manchon}(2022)}]{guemard:prb2022}%
  \BibitemOpen
  \bibfield  {author} {\bibinfo {author} {\bibfnamefont {V.}~\bibnamefont
  {Guemard}}\ and\ \bibinfo {author} {\bibfnamefont {A.}~\bibnamefont
  {Manchon}},\ }\href {\doibase 10.1103/PhysRevB.105.054433} {\bibfield
  {journal} {\bibinfo  {journal} {Phys. Rev. B}\ }\textbf {\bibinfo {volume}
  {105}},\ \bibinfo {pages} {054433} (\bibinfo {year} {2022})}\BibitemShut
  {NoStop}%
\bibitem [{\citenamefont {Rohling}\ and\ \citenamefont
  {Troncoso}(2023)}]{Rohling:prb2023}%
  \BibitemOpen
  \bibfield  {author} {\bibinfo {author} {\bibfnamefont {N.}~\bibnamefont
  {Rohling}}\ and\ \bibinfo {author} {\bibfnamefont {R.~E.}\ \bibnamefont
  {Troncoso}},\ }\href {\doibase 10.1103/PhysRevB.108.214416} {\bibfield
  {journal} {\bibinfo  {journal} {Phys. Rev. B}\ }\textbf {\bibinfo {volume}
  {108}},\ \bibinfo {pages} {214416} (\bibinfo {year} {2023})}\BibitemShut
  {NoStop}%
\bibitem [{\citenamefont {Tang}\ and\ \citenamefont
  {Bauer}(2023)}]{tang2023spin}%
  \BibitemOpen
  \bibfield  {author} {\bibinfo {author} {\bibfnamefont {P.}~\bibnamefont
  {Tang}}\ and\ \bibinfo {author} {\bibfnamefont {G.~E.}\ \bibnamefont
  {Bauer}},\ }\href {https://arxiv.org/abs/2312.09694} {\bibfield  {journal}
  {\bibinfo  {journal} {arXiv preprint arXiv:2312.09694}\ } (\bibinfo {year}
  {2023})}\BibitemShut {NoStop}%
\bibitem [{\citenamefont {Bender}\ \emph {et~al.}(2014)\citenamefont {Bender},
  \citenamefont {Duine}, \citenamefont {Brataas},\ and\ \citenamefont
  {Tserkovnyak}}]{bender2014dynamic}%
  \BibitemOpen
  \bibfield  {author} {\bibinfo {author} {\bibfnamefont {S.~A.}\ \bibnamefont
  {Bender}}, \bibinfo {author} {\bibfnamefont {R.~A.}\ \bibnamefont {Duine}},
  \bibinfo {author} {\bibfnamefont {A.}~\bibnamefont {Brataas}}, \ and\
  \bibinfo {author} {\bibfnamefont {Y.}~\bibnamefont {Tserkovnyak}},\ }\href
  {\doibase 10.1103/PhysRevB.90.094409} {\bibfield  {journal} {\bibinfo
  {journal} {Phys. Rev. B}\ }\textbf {\bibinfo {volume} {90}},\ \bibinfo
  {pages} {094409} (\bibinfo {year} {2014})}\BibitemShut {NoStop}%
\bibitem [{\citenamefont {Ashcroft}\ and\ \citenamefont
  {Mermin}(2022)}]{ashcroft2022solid}%
  \BibitemOpen
  \bibfield  {author} {\bibinfo {author} {\bibfnamefont {N.~W.}\ \bibnamefont
  {Ashcroft}}\ and\ \bibinfo {author} {\bibfnamefont {N.~D.}\ \bibnamefont
  {Mermin}},\ }\href@noop {} {\emph {\bibinfo {title} {Solid state physics}}}\
  (\bibinfo  {publisher} {Cengage Learning},\ \bibinfo {year}
  {2022})\BibitemShut {NoStop}%
\bibitem [{\citenamefont {Adachi}\ \emph {et~al.}(2013)\citenamefont {Adachi},
  \citenamefont {Uchida}, \citenamefont {Saitoh},\ and\ \citenamefont
  {Maekawa}}]{adachi2013theory}%
  \BibitemOpen
  \bibfield  {author} {\bibinfo {author} {\bibfnamefont {H.}~\bibnamefont
  {Adachi}}, \bibinfo {author} {\bibfnamefont {K.-i.}\ \bibnamefont {Uchida}},
  \bibinfo {author} {\bibfnamefont {E.}~\bibnamefont {Saitoh}}, \ and\ \bibinfo
  {author} {\bibfnamefont {S.}~\bibnamefont {Maekawa}},\ }\href {\doibase
  10.1088/0034-4885/76/3/036501} {\bibfield  {journal} {\bibinfo  {journal}
  {Reports on Progress in Physics}\ }\textbf {\bibinfo {volume} {76}},\
  \bibinfo {pages} {036501} (\bibinfo {year} {2013})}\BibitemShut {NoStop}%
\bibitem [{\citenamefont {Slonczewski}(1996)}]{slonczewski1996current}%
  \BibitemOpen
  \bibfield  {author} {\bibinfo {author} {\bibfnamefont {J.~C.}\ \bibnamefont
  {Slonczewski}},\ }\href {\doibase 10.1016/0304-8853(96)00062-5} {\bibfield
  {journal} {\bibinfo  {journal} {Journal of Magnetism and Magnetic Materials}\
  }\textbf {\bibinfo {volume} {159}},\ \bibinfo {pages} {L1} (\bibinfo {year}
  {1996})}\BibitemShut {NoStop}%
\bibitem [{\citenamefont {Tserkovnyak}\ \emph {et~al.}(2002)\citenamefont
  {Tserkovnyak}, \citenamefont {Brataas},\ and\ \citenamefont
  {Bauer}}]{tserkovnyak2002spin}%
  \BibitemOpen
  \bibfield  {author} {\bibinfo {author} {\bibfnamefont {Y.}~\bibnamefont
  {Tserkovnyak}}, \bibinfo {author} {\bibfnamefont {A.}~\bibnamefont
  {Brataas}}, \ and\ \bibinfo {author} {\bibfnamefont {G.~E.}\ \bibnamefont
  {Bauer}},\ }\href {\doibase 10.1103/PhysRevB.66.224403} {\bibfield  {journal}
  {\bibinfo  {journal} {Phys. Rev. B}\ }\textbf {\bibinfo {volume} {66}},\
  \bibinfo {pages} {224403} (\bibinfo {year} {2002})}\BibitemShut {NoStop}%
\bibitem [{\citenamefont {Flebus}(2019)}]{flebus2019chemical}%
  \BibitemOpen
  \bibfield  {author} {\bibinfo {author} {\bibfnamefont {B.}~\bibnamefont
  {Flebus}},\ }\href {\doibase 10.1103/PhysRevB.100.064410} {\bibfield
  {journal} {\bibinfo  {journal} {Phys. Rev. B}\ }\textbf {\bibinfo {volume}
  {100}},\ \bibinfo {pages} {064410} (\bibinfo {year} {2019})}\BibitemShut
  {NoStop}%
\end{thebibliography}
\end{document}